\documentclass[twocolumn,showpacs,amsmath,amssymb,superscriptaddress,aps,prc]{revtex4-1}
\usepackage{graphicx}
\usepackage{epstopdf}
\usepackage{epsfig}
\usepackage{bm}
\usepackage{color}
\begin{document}
\title{Inclusive Breakup Theory of Three-Body Halos}
%
%

\author{M. S. Hussein}
\affiliation{Instituto Tecnol\'ogico de Aeron\'autica, DCTA,
12228-900, S. Jos\'e dos Campos,~Brazil.}
\affiliation{Instituto de Estudos Avan\c{c}ados, Universidade de S\~{a}o Paulo C. P.
72012, 05508-970 S\~{a}o Paulo, SP, Brazil}
\affiliation{ Instituto de F\'{\i}sica,
Universidade de S\~{a}o Paulo, C. P. 66318, 05314-970 S\~{a}o Paulo, SP,
Brazil}
\author{L. A. Souza}
\affiliation{Instituto Tecnol\'ogico de Aeron\'autica, DCTA,
12228-900, S. Jos\'e dos Campos,~Brazil.}
\author{ E. Chimanski}
\affiliation{Instituto Tecnol\'ogico de Aeron\'autica, DCTA,
12228-900, S. Jos\'e dos Campos,~Brazil.}
\author{ B. V. Carlson}
\affiliation{Instituto Tecnol\'ogico de Aeron\'autica, DCTA,
12228-900, S. Jos\'e dos Campos,~Brazil.}
\author{T. Frederico} 
\affiliation{Instituto Tecnol\'ogico de Aeron\'autica, DCTA,
12228-900, S. Jos\'e dos Campos,~Brazil.}

%


\begin{abstract}
We present a recently developed theory for the 
inclusive breakup  of three-fragment projectiles within a four-body  spectator model \cite{CarPLB2017},  
 for the treatment of the elastic and 
inclusive non-elastic break up reactions involving weakly bound three-cluster nuclei in 
$A\,(a,b)\,X$ / $a = x_1 + x_2 + b$ collisions. The four-body theory is an extension
of  the three-body approaches developed in the 80's by 
Ichimura, Autern and Vincent (IAV) \cite{IAV1985}, Udagawa and Tamura (UT) \cite{UT1981} and Hussein and McVoy (HM) \cite{HM1985}. We expect that experimentalists shall be encouraged to search for more information about 
the $x_{1} + x_{2}$ system in the elastic breakup cross section and that also further developments  and 
extensions of the surrogate method will be pursued, based on the inclusive non-elastic breakup part of the $b$ spectrum.
\end{abstract}
\maketitle
\section{Introduction}
\label{intro}

We report a recently developed theory to treat the inclusive 
breakup of three-fragment weakly bound nuclei  \cite{CarPLB2017}. The natural three fragment 
candidates for projectiles are Borromean, two-nucleon and unstable three-fragment halo nuclei.
Our theory is an extension of  the inclusive breakup models used for  incomplete fusion reactions 
 and in the surrogate method with two fragment projectiles. The three-body approach was developed in the 80's by 
Ichimura, Autern and Vincent (IAV) \cite{IAV1985}, Udagawa and Tamura (UT) \cite{UT1981} and Hussein and McVoy (HM) \cite{HM1985}. These three-body theories were extended to obtain the fragment 
yield in the reaction $A\,(a,b)\,X$, where the projectile is $a = x_1 + x_2 + b$. 
The inclusive breakup cross section is a sum of the four-body 
elastic breakup cross section plus the inclusive non-elastic breakup cross section that involves the 
absorption cross sections of the participant fragments, $x_1$ and $x_2$, and which generalizes the three-body
formula reviewed in Austern, et al. \cite{Austern1987}. The connection between IAV, UT and HM theories was shown in \cite{HFM1990}.

The new  formula contains the four-body dynamics 
both in the elastic breakup cross section and in the inclusive 
non-elastic breakup ones. It can be applied to treat reactions with 
stable/unstable projectiles composed of three-fragments, like weakly bound Borromean and 
two-nucleon halo nuclei, where fingerprints of Efimov physics \cite{Efimov} 
and universality \cite{FrePPNP12}  could be revealed through the appearance of long-range correlations
between the three-fragments. In addition one can seek the generalization of the surrogate
method applied to reactions like (d,p) and (d,n) to the reactions (t,d) and ($^3$He,d), among others. 
Of particular interest is also 
the two-fragment correlation contribution to the elastic breakup cross section and the 
inclusive non-elastic breakup cross section through a three-body absorption interaction, which
appears naturally in the four-body formulation. One could examine, for example, the relation between 
the pairing, in the case of two-neutron halos, and the three-body formulation of the projectile and its
importance to the reaction mechanism underlying the inclusive breakup cross-sections.

We expect that these developments could stimulate experimentalists to search 
for more information about the $x_{1} + x_{2}$ system in the elastic breakup cross section, 
and that also further theoretical developments  and 
extensions of the surrogate method will be pursued, 
based on the inclusive non-elastic breakup part of the $b$ spectrum.

\section{Inclusive Break-up theory}
\label{sec-1}

 The many-body Hamiltonian for the $b + x_1 + x_2 + A$ system to be applied 
 to derive the scattering dynamics  of the three-fragment projectile is
\begin{multline}
 H_{(b, x_1, x_2, A)} = T_b + T_{{x}_1} + T_{{x}_2} + V_{b, {{x}_1}} + V_{b, {{x}_2}} + V_{{x}_1, {x}_2}
\\
 +  h_A + T_A +V_{b,A} +V_{{{x}_1},A} + V_{{{x}_2},A}   
\end{multline}
where the kinetic energies are given by the $T$'s  and the microscopic Hamiltonian of the target nucleus is 
$h_A$. In the  spectator approximation the 
microscopic potential $V_{b, A}$ is associated by standard methods in reaction theory  to the
optical potential $U_b$, and the target is considered infinitely massive, namely  $T_A$ = 0. 

The inclusive breakup cross-section is an integral over the spectator fragment position $\textbf{r}_b$ and a sum over  the $x_1 + x_2 + A$ bound and scattering states. This leads to the
 $b$-spectrum and angular distribution
\begin{multline}
\frac{d^{2}\sigma_b}{dE_{b}d\Omega_{b}} = \frac{2\pi}{\hbar v}\rho_{b}(E_b)\sum_{c}\,\delta(E - E_b -E_{c})\\ \times
 \left| \left\langle\chi_{b}^{(-)}\Psi^{c}_{x_{1}x_{2}A}\left|V_{b,x_{1}} + V_{b, x_2} + V_{x_1, x_2}\right|\varXi\right \rangle\right|^{2} \label{sigb1}
\end{multline}
where $\varXi(\textbf{r}_b, \textbf{r}_{{x_1}}, \textbf{r}_{{x_2}}, A)$ is the exact eigenstate of the $A + a$ many body 
Hamiltonian. The final state wave function is $\chi_{b}^{(-)}(\textbf{k}_{b}, \textbf{r}_{b})\Psi^{c}_{(x_{1}x_{2}A)}$,
where $\Psi^{c}_{(x_{1}x_{2}A)}$ runs over bound and continuum states. The  
 density of $b$ continuum states is $\rho_{b}(E_b) \equiv [d\textbf{k}_{b}/(2\pi)^3]/[dE_{b}d\Omega_{b}]  = \mu_{b}k_{b}/[(2\pi)^{3}\hbar^3]$, with
 $\mu_{b}$ the  reduced mass of the $b + A$ system.

The connection with the four-body (4B) scattering problem is obtained by eliminating the target internal degrees of freedom, 
and by using the product approximation, $\varXi = \Psi_0^{4B(+)}\Phi_{A}$, where
 $\Psi_{0}^{4B (+)}$ is the exact 4B
scattering wave function in the incident channel, and
 $\Phi_{A}$ is the ground state wave function of the target nucleus. Together with the 4B approximation, the
 energy conservation $\delta$ in Eq. (\ref{sigb1}) is associated with the
 imaginary part of an optical model Green's function operator
\begin{small}
\begin{equation}
\mbox{Im} G^{(+)}_X\, =\,
-\pi\,\Omega^{(-)}_X\,\delta(E_x - H_0)\,
(\Omega^{(-)}_X)^\dagger - 
(G^{(+)}_X)^\dagger\, W_X\,G^{(+)}_X 
\end{equation}\end{small}
where the M\"oller operator is $\Omega^{(-)}_{X} = [1 + G^{(-)}_{X}({V}_{X})^{\dagger}]$.
The imaginary part of the optical potential for particles $x_1$ and $x_2$ contains single fragment
terms, $W_{x_i}$, and a three-body term. In addition, closure has to be used 
to perform the sum over $c$ and  general nuclear reaction theory to transform 
 the microscopic interactions  $V_{x_1, A}$ and $V_{x_2, A}$ into complex optical potentials 
 $U_{x_1}$ and $U_{x_2}$ (see e.g. \cite{Austern1987}). 
 
 \subsection{Inclusive breakup cross-section}
 Following the above steps, the
inclusive breakup cross section is reduced to a sum of two distinct terms, 
the elastic breakup and the non-elastic breakup cross sections
\begin{equation}
\frac{d^{2}\sigma_b}{dE_{b}d\Omega_{b}} = \frac{d^{2}\sigma^{EB}_b}{dE_{b}d\Omega_{b}} + \frac{d^{2}\sigma^{INEB}_b}{dE_{b}d\Omega_{b}} \nonumber
\end{equation}
with the
 4B elastic breakup cross section contribution being
\begin{multline}
\frac{d^{2}\sigma^{EB}_b}{dE_{b}d\Omega_{b}} =  \frac{2\pi}{\hbar v_{a}}\rho_{b}(E_b)\int \frac{d{k}_{{x}_1}}{(2\pi)^{3}} \frac{d{k}_{{x}_2}}{(2\pi)^{3}}
\\
\times |\langle \chi^{3B(-)}_{{{x}_1}, {{x}_2}}\chi^{(-)}_{b}|V_{{bx_{1}}} + V_{{bx_{2}}} |\Psi_{0}^{4B(+)}\rangle|^2 
 \\ \times \delta(E - E_{b} - E_{({\textbf{k}_{{x}_1}}, \textbf{k}_{{x}_2})}) \label{s4bel}
\end{multline} 
 where $\chi^{3B(-)}_{{{x}_1}, {{x}_2}}$  is the full scattering wave function of the two unobserved fragments in the final channel.

 The inclusive cross-section for the inelastic breakup  
contains the optical potentials $U_{x_1}$, $U_{x_2}$ and the fragment-fragment interaction $V_{x_1, x_2}$ to all orders:
\begin{equation}
\frac{d^{2}\sigma^{INEB}_b}{dE_{b}d\Omega_{b}} = 
\frac{2}{\hbar v_a}\rho_{b}(E_b) \langle \hat{\rho}_{{{x}_1}, {{x}_2}}|{W_{x_1} + W_{x_2}} + { W_{3B}}|\hat{\rho}_{{{x}_1}, {{x}_2}}\rangle  \label{s4inel}
\end{equation}
where the source function
\begin{multline}
\hat{\rho}_{X}(\textbf{r}_{x_{1}},\textbf{r}_{x_{2}}) = (\chi_{b}^{(-)}|\Psi_{0}^{4B(+)}\rangle = \\ =
\int d\textbf{r}_{b}\left[\chi_{b}^{(-)}(\textbf{r}_{b})\right]^{\dagger}\Psi_{0}^{4B(+)}(\textbf{r}_{b}, \textbf{r}_{x_{1}}, \textbf{r}_{x_{2}})
\end{multline}
carries the full 4B dynamics in the optical model description. The
 imaginary parts of the optical potentials $U_{x_1}$ and $U_{x_2}$ are $W_{x_1}$ and $W_{x_2}$, respectively.
 Notice the presence of the imaginary part of a three-body optical potential ($W_{3B}$), associated with 
 inelastic excitations that are irreducible to single fragment inelastic processes. This will be discussed in more detail in a later sections. We point out
that the 4B inclusive non-elastic breakup cross section
differs significantly from the 3B Austern formula \cite{Austern1987}.

The inelastic breakup cross-section is a sum of three terms,
\begin{small}
\begin{equation}
\frac{d^{2}\sigma^{INEB}_b}{dE_{b}d\Omega_{b}} = \rho_{b}(E_b) \frac{k_a}{E_a}\left[\frac{{E_{x_1}}}{{k_{x_1}}}{ \sigma_{R}^{x_1}} + \frac{{E_{x_2}}}{{k_{x_2}}}{\sigma_{R}^{x_2}} + \frac{E_{CM}({{x}_1},{{ x}_2})}{(k_{{x}_1}+ k_{{x}_2})} { \sigma_{R}^{3B}}\right]
\end{equation}
\end{small}
where the approximation of a weakly bound projectile is used to write the kinetic energy:  $E_{{x_i},Lab} = E_{a, Lab}(M_{{x_{i}}}/M_a)$ with $M_{a}$ and $M_{{x_i}}$ being the  mass numbers of the projectile and fragment, respectively.
The single fragment  inclusive cross-sections are
\begin{small}
\begin{equation}
\sigma_{R}^{x_1} = \frac{{k_{x_1}}}{E_{{x_1}}} \langle \hat{\rho}_{{x}_1, {x}_2}|W_{x_1}|\hat{\rho}_{{x}_1, {x}_2}\rangle, \,\,\,\,\sigma_{R}^{x_2} = \frac{{k_{x_2}}}{E_{{x_2}}} \langle \hat{\rho}_{{x}_1, {x}_2}|W_{x_2}|\hat{\rho}_{{x}_1, {x}_2}\rangle, 
\end{equation}
\end{small}
and the double fragment inclusive cross-section is
\begin{equation}
\sigma_{R}^{3B} = \frac{(k_{{x}_1}+ k_{{x}_2})}{E_{CM}({{x}_1},{{ x}_2})}\ \langle \hat{\rho}_{{x}_1, {x}_2}|W_{3B}|\hat{\rho}_{{x}_1, {x}_2}\rangle\, \label{sig3b}
\end{equation}
which represents the two-fragment irreducible inelastic processes.
 
We elaborate here on the physical interpretation of the different contributions to the inclusive inelastic cross-section. 
The absorption of the fragment $x_i$ by the target is given by
$\sigma_{R}^{x_i}$ $(i=1,2)$, in the case in which th other fragment 
$x_j$ just scatters off the target through the optical potential $U_{{x}_j A}$. It is noteworthy to 
stress that $\sigma_{R}^{x_i}$ is different from the one in the 3B theory of the
$b - x -A$ system. The cross-sections
$\sigma_{R}^{x_1}$, $\sigma_{R}^{x_2}$ are related to the three-body IAV cross section 
through a convolution of the latter with the distorted wave densities  $|\chi^{(+)}_{{x}_2}(\textbf{r}_{x_2})|^2$, 
and $|\chi^{(+)}_{{x}_1}(\textbf{r}_{x_1})|^2$ of the spectator fragments $x_1$ and $x_2$, respectively. 
If an eikonal-type approximation of the projectile distorted wave $\chi_{a}^{(+)} \times \Phi_a({r_b},{{{r_{x_{1}}}}}, {{{r_{x_{2}}}}})$ ($a\equiv x_1+x_2+b$) is used in the 4B theory, the difference  with the 3B formulation is 
clearly exposed. Thus, it is expected that the
 $\sigma_{R}^{{x}_1}$ in a (t,p) reaction will differ from the $\sigma_{R}^{x}$ extracted in a (d,p) reaction. 
Furthermore, at low energies  $\sigma_{R}^{x_1}$, $\sigma_{R}^{x_2}$, and $\sigma_{R}^{3B}$ are associated with the 
formation of compound nuclei  $A +x_1$, $A+x_2$ and $A + (x_1 +x_2)$.

\subsection{Examples}
\label{sec-3}

The first example to which we apply  the 4B formulation of the inelastic breakup cross-sections is the case of the reaction $^9$Be (= $\alpha + \alpha + n$) + $^{208}$Pb, where
4B-CDCC (four-body Continuum Discretized Coupled Channels) calculations were performed by 
Descouvemont and collaborators \cite{DDCH2015} for the elastic scattering,  elastic breakup, and 
total fusion cross-sections. However so far, CDCC cannot obtain the partial or incomplete fusion 
cross-sections. Despite this fact, we can point out that compound nuclei formation 
with $\alpha$ detected, namely $\alpha$ + $^{208}$Pb = $^{210}$Po, n + $^{208}$Pb = $^{209}$Pb, and $\alpha$ + n + $^{208}$Pb = $^{211}$Po, would be very interesting to observe, in order to gather information about the spectrum of $\alpha$ particles in the analysis of the inclusive cross sections. It would be interesting to investigate the properties of these compound nuclei experimentally, as they are formed in such a hybrid reaction.

A second example is a two-neutron Borromean nuclei. The inclusive $\alpha$ spectrum for the $^6$He projectile
accounts for the 
 formation of n + A and 2n + A compound nuclei. However it would be experimentally difficult to distinguish these in CN decay. 

The third example is a two-proton  Borromean nuclei.  Onemight consider the inclusive proton spectra in the breakup 
of $^{20}$Mg + $^{208}$Pb with the formation of  $^{209}$Bi, $^{226}$U and  $^{227}$Np at different excitation energies.
However, $^{20}$Mg presents a very short lifetime and  low intensity as a   secondary beam.

\section{ IB Cross Sections: HM Source function  }

The source function computed with the  Hussein and McVoy (HM) model \cite{HM1985} can be written as:
\begin{equation}
\langle \textbf{r}_{{x}_1}, \textbf{r}_{{x}_2}|\hat{\rho}^{4B}_{HM}\rangle = \hat{S}_{b}(\textbf{r}_{{x}_1}, \textbf{r}_{{x}_2})\chi^{(+)}_{{x}_1}(\textbf{r}_{{x}_1})\chi^{(+)}_{{x}_2}(\textbf{r}_{{x}_2})
\end{equation}
where the four-body scattering state is approximated by the product of the distorted waves of the three
fragments and the projectile bound state wave function. The 
 internal motion modifies the S-matrix of the b fragment as
\begin{equation}
 \hat{S}_{b}(\textbf{r}_{{x}_1}, \textbf{r}_{{x}_2}) \equiv \int d\textbf{r}_{b} { \Phi_{a}(\textbf{r}_{{x}_1}, \textbf{r}_{{x}_2}, \textbf{r}_b)} { \langle\chi^{(-)}_{b}|\chi^{(+)}_{b}\rangle(\textbf{r}_b)} \, ,
 \end{equation}
which should be compared to the 3B
 S-matrix element  of $b$ given by  $S_{\textbf{k}_{b}^{\prime}, \textbf{k}_{b}}=\int d\textbf{r}_{b} \langle\chi^{(-)}_{b}|\chi^{(+)}_{b}\rangle(\textbf{r}_b)$. In this approximation the
 inclusive cross-section in the 4B theory with $a=x_1+x_2+b$ is
 \begin{multline}
\frac{E_{{x}_1}}{k_{{x}_1}}\sigma_{R}^{x_1} = \int d\textbf{r}_{{x}_1} \int d\textbf{r}_{{x}_2} |\hat{S}_{b}(\textbf{r}_{{x}_1}, \textbf{r}_{{x}_2})|^{2} \\ \times |\chi^{(+)}_{{x}_2}(\textbf{r}_{x_2})|^2
\, W(\textbf{r}_{{x}_1})|\chi^{(+)}_{{x}_1}(\textbf{r}_{x_1})|^2 \, , \label{s4b}
\end{multline}
which should be contrasted with the cross-section in the 3B theory $(a=x+b)$
\begin{equation}
\frac{E_{x}}{k_x}\sigma_{R}^{x} = \int d\textbf{r}_{x} { |\hat{S}_{b}(\textbf{r}_x)|^{2}} W(\textbf{r}_x)|\chi^{(+)}_{x}(\textbf{r}_{x})|^2 \, , \label{s3b}
\end{equation}
where $ \hat{S}_{b}(\textbf{r}_x) \equiv \int d\textbf{r}_{b}  \langle\chi^{(-)}_{b}|\chi^{(+)}_{b}\rangle(\textbf{r}_b)\Phi_{a}(\textbf{r}_b, \textbf{r}_x)$. Comparing the cross-sections
in Eqs. (\ref{s4b}) and (\ref{s3b}), one singles out the distorted wave 
$|\chi^{(+)}_{{x}_2}(\textbf{r}_{x_2})|^2$ , which damps the 4B cross-section with respect to the
3B one.

\section{HM and 3B Glauber theory }

\begin{figure}[htb]
   \centerline{\epsfig{figure=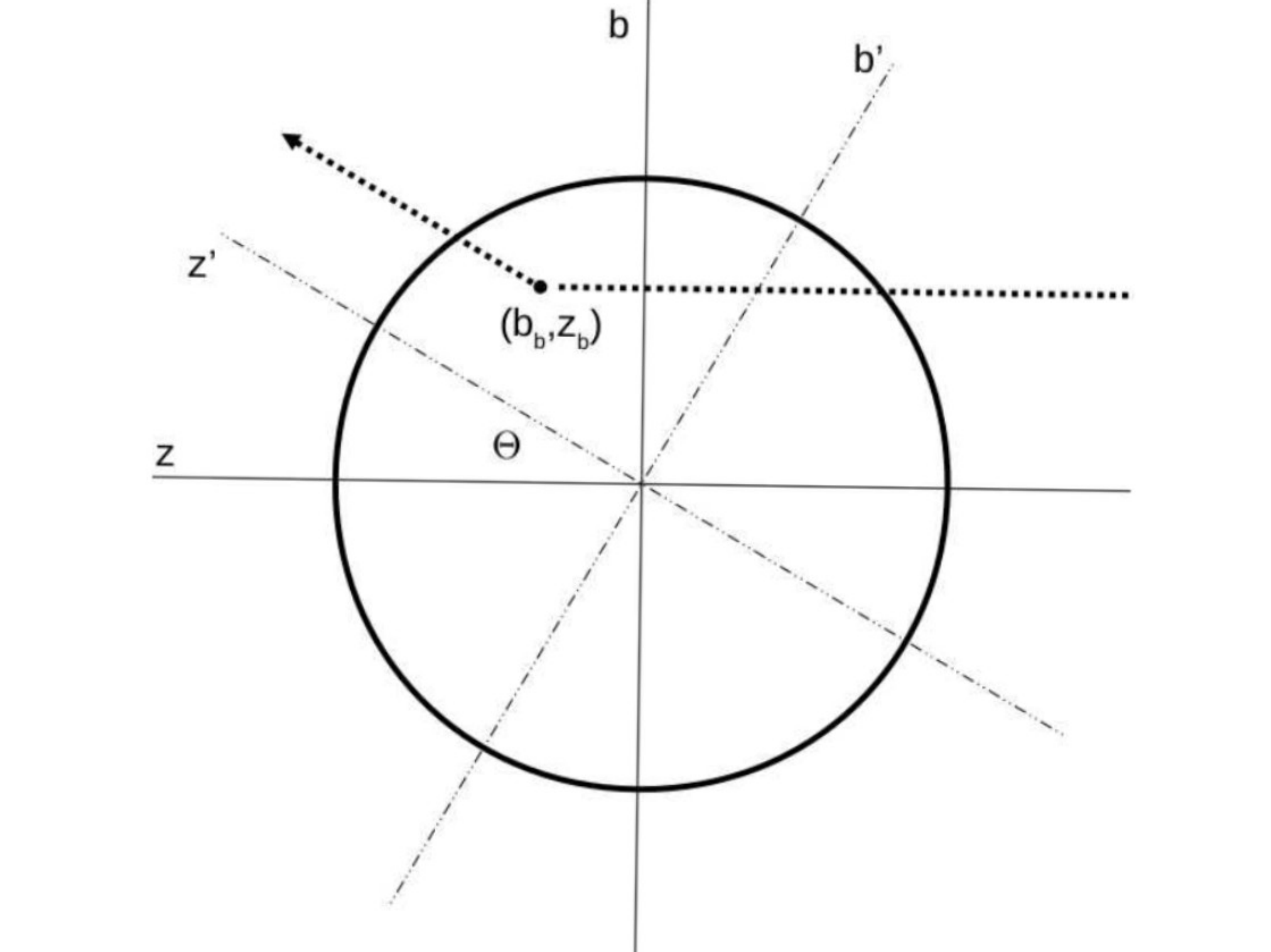,height=6cm}}
\caption{ Appropriate kinematics and definitions of reference frame to compute the Glauber formula for the spectator distorted waves.}
\label{fig1}
\end{figure}

We begin with the Glauber phase \cite{CH2013}
\begin{small}
\begin{multline}
[\psi_{\vec{p}_{b}^{\prime}}^{(-)}\left(\vec{r^{\prime}}_{b}\right)]^{\star}\psi_{\vec{p}_{b}}^{(+)}\left(\vec{r}_{b}\right)
=\exp\left[-i\frac{m}{\hbar p_{b}^{\prime}}\int_{-\infty}^{z_{b}^{\prime}}dz^{\prime}V\left(\sqrt{b_{b}^{\prime2}+z^{\prime2}}\right)
\right.\\ \left.
-i\frac{m}{\hbar p_{b}}\int_{z_{b}}^{\infty}dz^{\prime}V\left(\sqrt{b_{b}^{2}+z^{\prime2}}\right)
-\imath\, \vec q\cdot \vec r_b\right]\,,
\end{multline}
\end{small}
where $\hbar \vec q= {\vec p}_b^{\,\prime} -\vec p_b$, and $\vec p_b$ (${\vec p}^{\,\prime}_b$)  is the incoming 
(outgoing) momentum of the spectator. We take the initial value of the momentum to be $\vec{p}_{b}=p_{b}\hat{k}$
and the final value to be $\vec{p}_{b}^{\,\prime}=p_{b}^{\prime}\sin\theta\,\hat{i}+p_{b}^{\prime}\cos\theta\,\hat{k}$.
The coordinates $(b_{b}^{\prime},\,z_{b}^{\prime})$ can be written
in terms of the coordinates $(b_{b},\,z_{b})$ as 
\begin{eqnarray}
b_{b}^{\prime} & = & b_{b}\cos\theta-z_{b}\sin\theta\\
z_{b}^{\prime} & = & b_{b}\sin\theta+z_{b}\cos\theta\,.
\end{eqnarray}
Figure \ref{fig1} illustrates the coordinates $(b_{b},\,z_{b})$ and $(b_{b}^{\prime},\,z_{b}^{\prime})$.

In order to get insight into the structure of the source function, we approximate the potential by a square well 
\begin{equation}
V\left(r\right)=\left(V_{0}\,-\, \imath\, W_{0}\right)\Theta\left(R-r\right)\,.
\end{equation}
The spectator distorted  density then becomes
\begin{small}
\begin{multline}
[\psi_{\vec{p}_{b}^{\prime}}^{(-)}\left(\vec{r^{\prime}}_{b}\right)]^{\star}\psi_{\vec{p}_{b}}^{(+)}\left(\vec{r}_{b}\right)=
\exp\left[-i\frac{m}{\hbar^{2}}\left(V_{0}+iW_{0}\right)\right.
\\ \left. \times
\left(\frac{1}{p_{b}^{\prime}}\left(\sqrt{R^{2}-b_{b}^{\prime2}}-z_{b}^{\prime}\right)+\frac{1}{p_{b}}\left(z_{b}+\sqrt{R^{2}-b_{b}^{2}}\right)\right)
-\imath\, \vec q\cdot \vec r_b
\right] 
\end{multline}
\end{small}
for  $b_{b}<R$ and $\exp[-\imath\, \vec q\cdot \vec r_b]$ for $b_{b}\ge R$.

The formula simplifies for the case of strong absorption, where the spectator distorted wave density becomes
\begin{equation}
[\psi_{\vec{p}_{b}^{\prime}}^{(-)}\left(\vec{r^{\prime}}_{b}\right)]^\star\psi_{\vec{p}_{b}}^{(+)}\left(\vec{r}_{b}\right)
=\Theta\left(b_{b}-R\right)\,\Theta\left(b_{b}^{\prime}-R\right)
\,\text{e}^{-\imath\, \vec q\cdot \vec r_b}\,.
\end{equation}
Standard eikonal calculations ignore the difference $(b_{b},\,z_{b})\,\neq\,(b_{b}^{\prime},\,z_{b}^{\prime})$, inwhich case the distorted wave density
can be simply written as
\begin{equation}
[\psi_{\vec{p}_{b}^{\prime}}^{(-)}\left(\vec{r^{\prime}}_{b}\right)]^{\star}\psi_{\vec{p}_{b}}^{(+)}\left(\vec{r}_{b}\right)
=\Theta\left(b_{b}-R\right)\,\text{e}^{-\imath\, \vec q\cdot \vec r_b}
\end{equation}

\begin{figure}[htb]
   \centerline{\epsfig{figure=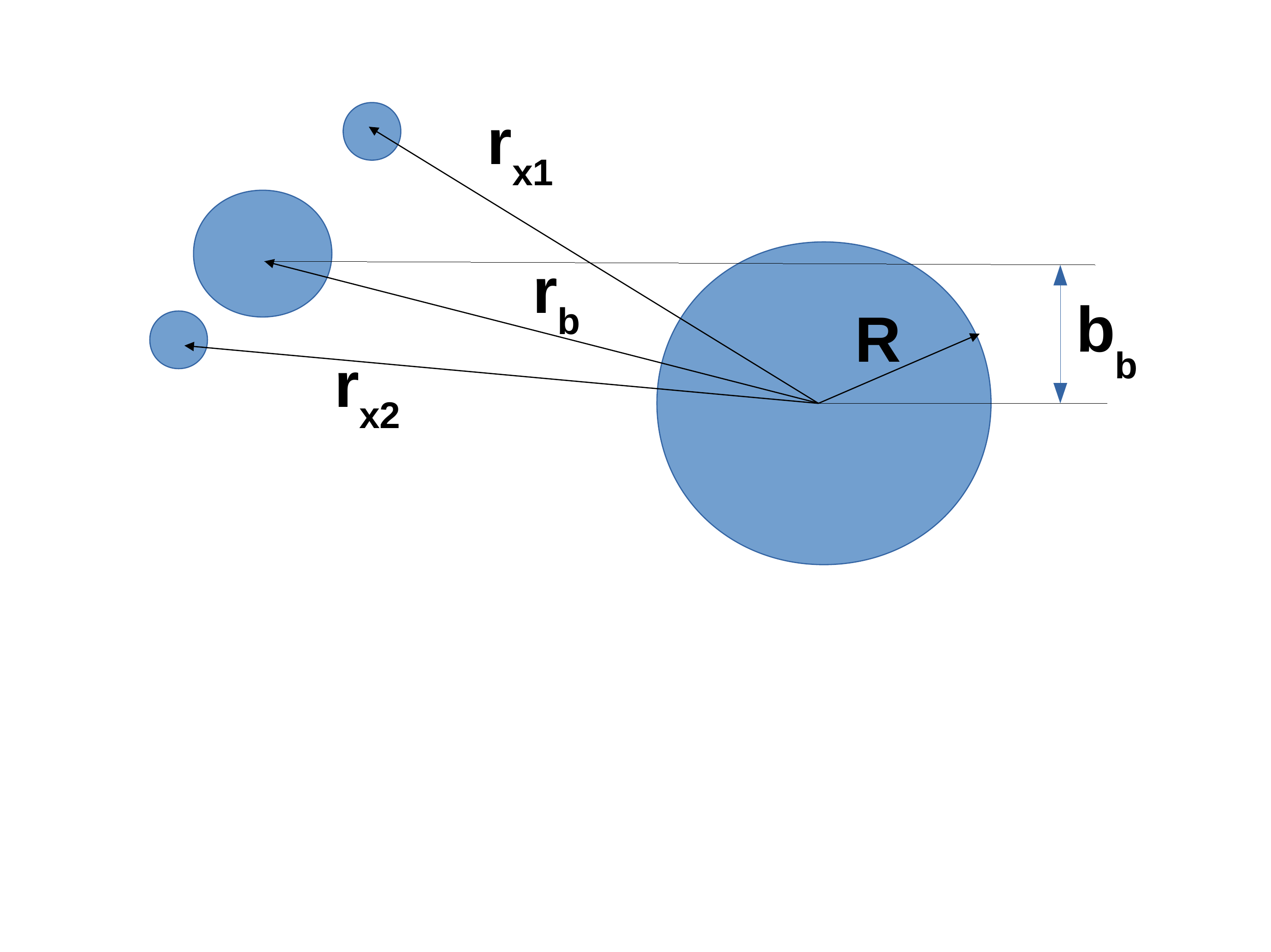,height=6cm}}
\vspace{-2cm}
\caption{Definition of the coordinates of the fragments in configuration space. The target is represented by the largest circle, the unobserved fragments $x_1$ and $x_2$ are represented by the smallest circles, and the remaining circle represents the detected particle $b$.}
\label{fig2}
\end{figure}

The internal motion modified S-matrix of the b fragment within the approximations above is given by
\begin{equation}
 \hat{S}_{b}(\textbf{r}_{{x}_1}, \textbf{r}_{{x}_2}) = \int d\textbf{r}_{b} { \Phi_{a}(\textbf{r}_{{x}_1}, \textbf{r}_{{x}_2}, \textbf{r}_b)}  \Theta\left(b_{b} -R\right)\,\text{e}^{-\imath\, \vec q\cdot \vec r_b} \label{s4b1}
 \end{equation}
which produces a long range correlation between the fragments  $x_1$ and $x_2$ for weakly bound projectiles, due to
the long tail of the wave function penetrating in the classically forbidden region. This give us a
taste of the  possible r\"{o}le of Efimov physics, for example when the projectile is a 
two-neutron halo s-wave state such as $^{11}$Li, in creating a long range correlation between the two neutrons. The 
actual set of coordinates to compute (\ref{s4b1}) can be visualized  in Fig. \ref{fig2}.

 Figure \ref{fig2} also supplies a glimpse of the effect of an extended weakly bound system on the correlation between the two fragments $x_1$ and $x_2$. 
The formula (\ref{s4b1}) restricts $b$ to be  outside the target absorptive region. However the 
tail of the projectile wave function can provide a long range correlation between the two fragments and the target. 

It is instructive to calculate $\hat{S}_{b}(\textbf{r}_{{x}_1}, \textbf{r}_{{x}_2})$ in the case of two-fragment projectiles, such as the deuteron in the $(d, p)$ reaction. Here $b = p$ and $x = n$, and we obtain, within the same strong absorption (black disk)  eikonal approximation,
\begin{equation}
\hat{S}_{p}(\textbf{r}_{n}) = \int d\textbf{r}_{p} { \Phi_{d}(\textbf{r}_{n}, \textbf{r}_p)}  \Theta\left(b_{p} -R\right)\,\text{e}^{-\imath\, \vec q\cdot \vec r_p} \label{s4b2}
\end{equation}
which corresponds to an incomplete Fourier transform involving $\textbf{r}_p$ and $q$.

In the high energy regime, the cross section can be represented as an integral over impact parameter. Since Eq. (\ref{s3b}),
\begin{equation}
\frac{E_{x}}{k_x}\sigma_{R}^{x} = \int d\textbf{r}_{x}  |\hat{S}_{b}(\textbf{r}_x)|^{2} \left[W(\textbf{r}_x)|\chi^{(+)}_{x}(\textbf{r}_{x})|^2\right] ,
\end{equation}
contains the integrand of a b-integral of the reaction cross section of $x$, $[W(\textbf{r}_x)|\chi^{(+)}_{x}(\textbf{r}_{x})|^2]$,  which can be replaced by $[1 - |S_{x}(b_x)|^2]$, and the 
factor $|\hat{S}_{b}(\textbf{r}_x)|^2$ is basically the survival probability of the observed fragment, $b$, one can write, after rearranging terms,
\begin{equation}
\sigma_{R}^{x} = 2\pi \int db \langle\Phi_{a}||S_{b}(b)|^2\left[1 - |S_{x}(b)|^2\right]\Phi_{a}\rangle
\end{equation}
One can say that the above equation is the high energy eikonal limit of IAV or the HM formulae. At high energy, the source function Eq.(\ref{rhosum}) for two-fragment projectiles reduces to $\hat{\rho}^{IAV}_{x} = \hat{\rho}^{HM}_{x} = \text{(}\chi^{(-)}_{b}|\chi^{(+)}_{b}\chi^{(+)}_{x}\Phi_{a}\rangle = \langle\chi^{(-)}_{b}|\chi^{(+)}_{b}\rangle  |\chi^{(+)}_{x}\Phi_{a}\rangle$. When used in the expression for the cross section Eq. (\ref{s3b}) or in its original form,
\begin{equation}
\frac{d^{2}\sigma^{INEB}_b}{dE_{b}d\Omega_{b}} = 
\frac{2}{\hbar v_a}\rho_{b}(E_b) \langle \hat{\rho}^{HM}_{x}|{W_{x}|\hat{\rho}^{HM}_{x}}\rangle\,,  \label{s3inel}
\end{equation}
it is a simple exercise to reduce the expression to the Glauber cross section above. The above form of the cross section has been extensively used by \cite{Bertsch, Tostevin} to calculate one nucleon stripping and pickup reactions. These authors, take $\hat{R}_1 = |\hat{S}_{b}(\textbf{r}_x)|^{2} \left[W(\textbf{r}_x)|\chi^{(+)}_{x}(\textbf{r}_{x})|^2\right]$ as an operator and calculate the expectation value in the state from which the nucleon is removed or added, $\langle \Phi_{i}|\hat{R}_1|\Phi_{i}\rangle$. This is then used to represent the cross section in a particularly transparent form. The partial cross section for removal of a nucleon, from a single-particle configuration $j^{\pi}$ populating the residue final state 
$\alpha$ with excitation energy $E^{\star}_{\alpha}$, is calculated as

\begin{equation}
\sigma_{\alpha} = \left(\frac{A}{A - 1}\right)^{N}C^{2}S(\alpha, j^{\pi}) \langle\Phi_{j^{\pi}}|\hat{R}_1|\Phi_{j^{\pi}}\rangle 
\end{equation}

where $S_{\alpha}^{\star}= S_{n,p} + E_{\alpha}^{\star}$ is the effective separation energy for the final state $\alpha$ and $S_{n,p}$ is the ground-state to
ground-state nucleon separation energy. Here the factor $N$, in the 
A-dependent center-of-mass correction factor that multi-
plies the shell-model spectroscopic factors $C^{2}S(\alpha,j^{\pi})$, is
the number of oscillator quanta associated with the major
shell of the removed particle \cite{Tostevin}. The cross section for the removal of a nucleon from the $j^{\pi}$ single  particle shell is then the sum over all bound final states, $\alpha$ of the residual nucleus $A -1$

Here, the radioactive projectile is now denoted by A. In this manner, useful nuclear structure information contained in the spectroscopic factor of the fragment in the projectile A, $C^{2}S(\alpha, j^{\pi})$, can be extracted. Further, within the four-body theory, the practitioners of the eikonal theory extend their application to two nucleon removal or pickup. They use the following form of the reaction factor $\langle\Phi_{i}|\hat{R}_2|\Phi_{i}\rangle = \langle\Phi_{i}||S_{b}|^{2}| (1 - |S_{x_{1}}|^2)(1 - |S_{x_{2}}|^2)|\Phi_{i}\rangle$. This formula should come out from $\sigma_{R}^{3B}$ of Eq. (\ref{sig3b}) of our theory in the high energy limit. However, there are several shortcomings in the above model, as there is no reference to the individual absorptions of $x_1$ and $x_2$. Further, the model above misses the correlation between the two interacting fragments. Currently we are investigating this point.

\section{Double fragment inelastic processes }

The inclusive cross-section expressed by $\sigma_{R}^{3B}$ given by Eq. (\ref{sig3b}) is new and corresponds to
 genuine three-body absorption processes to inelastic channels. This part of the cross-section is associated with
 a  three-body optical potential $U_{3B}$ depending  on the relative coordinates of the fragments $x_1$, $x_2$ 
 and the target, and cannot be reduced to connected terms of the 3B transition matrix with two-body 
 potentials in different subsystems and the target in the ground state.

 \begin{figure}[htb]
  \centerline{\epsfig{figure=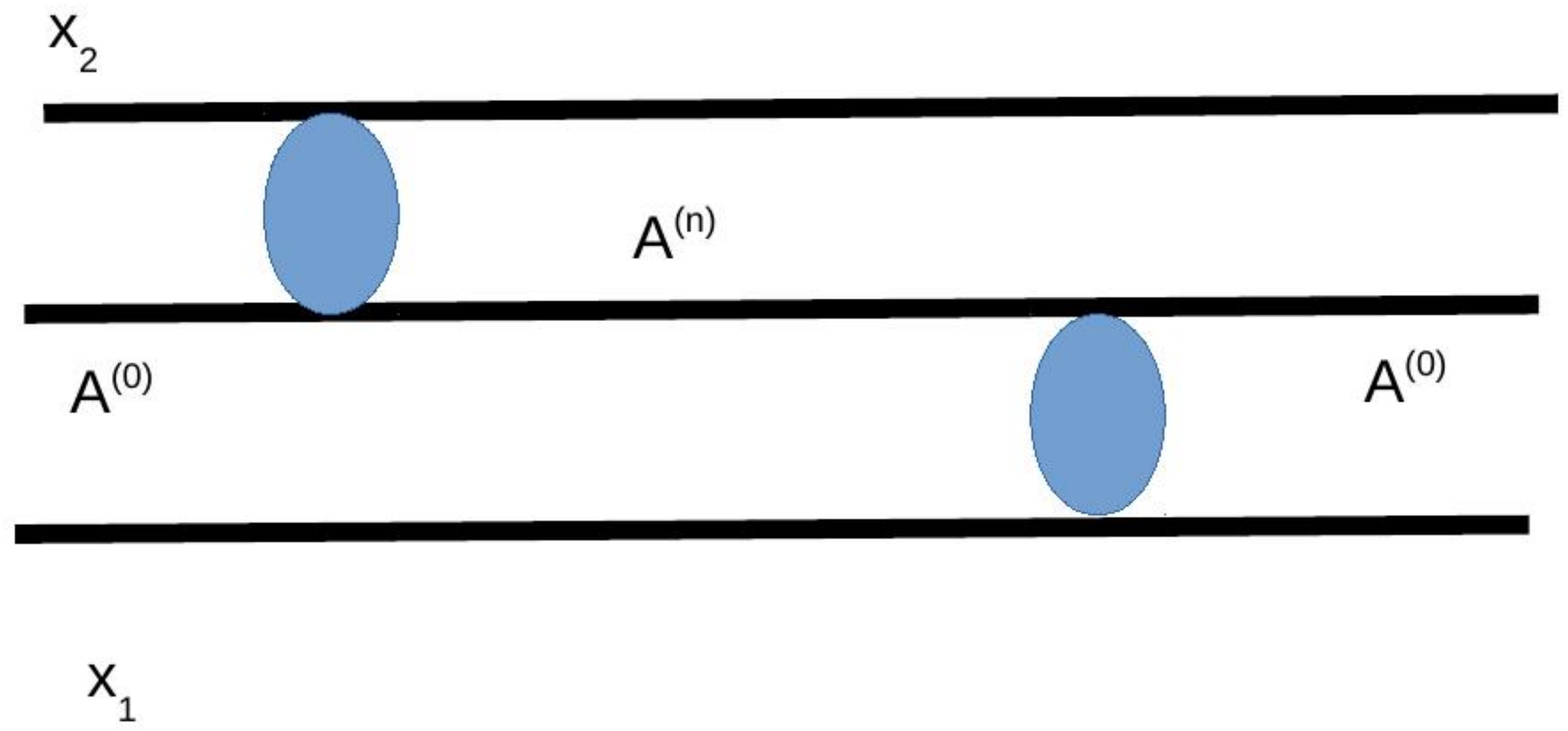,height=6cm}}
\vspace{-1cm}
\caption{ Three-body optical potential $U_{3B}$. Excitation of the target by particle $x_1$ and 
de-excitation by $x_2$.}
\label{fig3}
\end{figure}
\vspace{-0.5cm}

The structure of $W_{3B}=\mbox{Im}[U_{3B}]$ results from conventional nuclear reaction theory. It
includes, for example, processes like the 
virtual excitation of the target by one of the fragments and its virtual de-excitation by the other, as well as 
other 3B processes irreducible to the re-scattering terms with the target in the ground state and 
with the final result being the full capture, or complete fusion, of both fragments. This is illustrated
schematically in 
figure \ref{fig3}. If a resonant process exists, such as the excitation of giant pairing vibrations from the 
transfer of two neutrons from the projectile, it should furnish a large contribution to the optical 
three-body potential at the resonance energy.

\subsection{Hint on the 3B optical potential $U_{3B}$ }

The schematic figure \ref{fig3} furnishes a path to formally  build the
the three-body optical potential by using projection operators out 
of the target ground state,  
\begin{multline}
U_{3B} = PV_{{x_1}A}Q (QG_{xA}(E_{x})Q) QV_{{x_2}A}P \\ + 
PV_{{{x}_2}A}Q (QG_{xA}(E_x)Q) QV_{{{x}_1}A}P\,, \label{u3b}
\end{multline}
where $E_x=E_{x_1}+E_{x_2}$ and the Q-projected 3B Green's function of the $x_1 + x_2 + A$ 
system is
\begin{equation}
QG_{xA}Q = 
\left[E_{x} - QH_{0}Q + Q\, V_{xA}\,P\,G_{0}P\, V_{xA}\,Q + i\varepsilon\right]^{-1}
\end{equation}
with $V_{xA}\equiv V_{{{x}_1}A} + V_{{{x}_2}A}$. 
The imaginary part of the Q-projected Green's function of the ${{x}_1}{{x}_2}A\equiv xA$ 
system includes the virtual propagation n all states except for the elastic channel,
\begin{multline}
Im[QG_{xA}Q] = -\pi \Omega^{(-)}_{Q}Q\delta(E_x - H_{0})Q(\Omega^{(-)}_{Q})^{\dagger} \\ +
(QG_{xA}Q)^{\dagger} 
Q\, V_{xA}\,P\delta(E_x- H_{0})P\, V_{xA}\, QG_{xA}Q   
\end{multline}
Then, again using standard 
techniques, the imaginary part of $U_{3B}$ can be isolated in (\ref{u3b}) and can be written as 
 \begin{multline}
W_{3B}=U_{3B}^\dagger-U_{3B} =\\=  \pi[PV_{{x_1A}}Q \Omega^{(-)}_{Q}Q\delta(E_x - H_{0})Q 
(\Omega^{(-)}_{Q})^{\dagger}QV_{{x_2A}}P  
\\
 + PV_{{x_1A}}Q (QG_{xA}Q)^{\dagger} Q\,V_{{xA}}\,
P\,\delta(E_x - H_{0})\,P \\ \times  P\,V_{{xA}}\,Q  (QG_{xA}Q) QV_{{xA}}P]
 +(x_{1}\leftrightarrow x_{2})
\end{multline}
where one can identify the virtual propagation of the $x_1x_2A$ system through the inelastic channels
in the Q-space.

\section{Beyond HM, UT, and IAV formulas}

The key points for going beyond the IAV formula for the inclusive inelastic cross-section 
are to include in the source term the full four-body dynamics and the three-body optical potential,
by computing $|\Psi^{4B(+)}_0\rangle$ without approximating the four-body problem.
For the time being we will be content to generalize the three-body IAV formula to the 
four-body case and present the CFH formula given in \cite{CarPLB2017}. 

The  generalization given by CFH of the IAV formula to determine the inclusive
cross-section observed by the detection of the spectator particle $b$, contains
HM and UT like terms. The steps to derive a 4B version of the IAV formula follow
 \cite{HFM1990} and we write
 \begin{small}
\begin{equation}
|\Psi^{4B(+)}_0\rangle=G^{(+)}_{b, x_1, x_2, A}V_x|\Psi^{4B(+)}_0\rangle
\approx\, G^{(+)}_{b, x_1, x_2, A}\,V_x\, |\chi^{(+)}_a\,,\,\Phi_a\rangle 
\end{equation}
\end{small}
where $V_x=V_{b\,x_1}+V_{b\,x_2}$ is the interaction between the detected fragment and the other two, 
$|\Phi_a\rangle$ the projectile internal wave function and $\chi^{(+)}_a$ the distorted wave of the 
center of mass. The four body Green's function is
\begin{equation} 
 G^{(+)}_{b, x_1, x_2, A} =
[E - T_{b} - T_{x_1} - T_{x_2} - V_{x_1, x_2} - U_{b} - U_{x_1} - U_{x_2} + i\varepsilon]^{-1}.
\end{equation} 
where we have not explicitly included $U_{3B}$, which should also contribute to the Green's function.

The generalization of the IAV three-body source function for the four-body case is:
\begin{equation}
\hat{\rho}^{CFH}_{x_1,x_2} =(\chi_{b}^{(-)}|\Psi_{0}^{4B(+)}\rangle = 
\langle\chi_{b}^{(-)}|G^{(+)}_{b, x_1, x_2, A}\,V_{x}\,|\chi^{(+)}_{a}\Phi_{a}\rangle 
\label{CFH}
\end{equation}
By manipulating the above formula in analogy to the development  provided in
Ref. \cite{HFM1990}, one gets the 4B extension of the 3B Ichimura-Austern-Vincent as
\begin{equation}
\hat{\rho}_{x_1, x_2}^{CFH} =
\hat{\rho}_{x_1, x_2}^{UT} +\hat{\rho}_{x_1, x_2}^{HM}   \label{rhosum}
\end{equation}
where the 4B Hussein-McVoy (HM) term is
\begin{equation}
\hat{\rho}_{x_1, x_2}^{HM} = \langle\chi_{b}^{(-)}|\chi_{a}^{(+)}\Phi_{a}\rangle
\end{equation}
and the 4B Udagawa-Tamura (UT) term is 
\begin{equation}
\hat{\rho}_{x_1, x_2}^{UT} \equiv G^{(+)}_{x_1, x_2, A}\langle\chi_{b}^{(-)}|\left[U_b + U_{x_1} + U_{x_2} - U_a\right]|\chi^{(+)}_{a}\Phi_{a}\rangle  \,.
\end{equation}
The Green's function 
\begin{equation}
G^{(+)}_{x_1, x_2, A} = [E - E_b - T_{x_1} - T_{x_2} - V_{x_1, x_2} - U_{x_1} - U_{x_2} + i\varepsilon]^{-1}
\end{equation}
should also contain in principle the three-body optical potential. Further studies of this particular point 
will be developed.

\section{ Inclusive breakup and Efimov Physics }

The inclusive cross-section formulae for the elastic breakup (\ref{s4bel}) and 
for the inelastic process (\ref{s4inel})  applied to a three-body halo nuclei, e.g., a weakly bound 
neutron-neutron-core nuclei close to the drip-line, contains a long-range correlation between the two neutrons and the
core $b$, through the extended wave function of the halo and the scattering wave 
$\chi^{(+)}_{{x}_1}(\textbf{r}_{x_1})\chi^{(+)}_{{x}_2}(\textbf{r}_{x_2})\chi^{(+)}_{{x}_b}(\textbf{r}_{x_b})$, which
appears both in (\ref{s4bel}) and in the source function in (\ref{s4inel}), when such an approximation is made.

However, we expect that interesting physics remains beyond such an approximation if  the 4B scattering wave function, 
$\Psi_{0}^{4B(+)}(\textbf{r}_{b}, \textbf{r}_{x_{1}}, \textbf{r}_{x_{2}})$ is taken in full. 
It should contain dynamics of the continuum of the neutron-neutron-core system beyond the bound halo state. The reaction 
mechanism now includes Efimov physics \cite{Efimov} and with that the long range correlation between the halo fragments, which could be influenced by the presence of Efimov bound, virtual or resonant states \cite{FrePPNP12} (see also \cite{MaccFBS15} for a discussion
of observing Efimov states in halo reactions). Indeed, the existence of Borromean Efimov states was observed about a
decade ago \cite{InnsNP06} in the resonant three-body recombination of cold cesium atoms in magneto-optical traps.

\section{Digression on two-fragment projectile inclusive breakup cross section}

It is important to remind the reader of the advances made in the application of the inclusive breakup theory of two-fragment projectiles, as was originally developed in \cite{IAV1985, UT1981, HM1985, Austern1987}. Quite recently this theory was applied to the (d,p) reaction \cite{Ducasse2015, Potel2015, Moro2015, Carlson2015} as a mean to test the validity of the Surrogate Method \cite{SM0, SM1, SM2}, employed to extract neutron capture and fission cross sections of actinide target nuclei, such as $^{238}$U, $^{232}$Th, of importance for the development of next generation fast breeder reactors. These neutron capture reactions on other targets are also important for the study of element formation following supernova explosion through the astrophysical s-process. The general conclusion of  \cite{Ducasse2015, Potel2015, Moro2015, Carlson2015} was that the Surrogate Method, the extraction of the cross section for the formation and decay of the compound nucleus of the $(n + A)$ subsystem, as well as the total capture cross section (the cross section for the formation of the compound nucleus),  is justified as long as the direct part of the cross section is calculated and subtracted from the inclusive breakup total reaction cross section $\sigma^{(n + A)}_{R}$. Another topic of importance is the Trojan Horse Method \cite{THM0, THM1, THM2, THM3} used to extract from the inclusive breakup data a direct reaction of interest to nuclear astrophysics following nova explosion, which otherwise would be difficult to measure in the laboratory. The THM uses the advantage that as the surrogate charged fragment $x$ is brought by the primary projectile to the region of the target, the Coulomb barrier is already surmounted, and accordingly no hinderance due to barrier penetration and tunneling is present in the $x$ induced reaction. Accoringly, the x-induced reaction proceeds above the Coulomb barrier. Being so, the electron screening problem is also avoided.\\ 

The two methods, SM, and THM, are two pieces of the same quantity, namely $\sigma^{(x + A)}_{R}$ in the breakup reaction, $a + A \rightarrow b + (x + A)$. Thus the THM can be easily justified within the theory alluded to above as being a process contributing to the direct part of $\sigma^{(x + A)}_{R}$. As a final remark, if the surrogate fragment $x$ is a neutron then the SM and the THM are the same!

It would certainly be important to extend these studies to the case of the three-fragment projectiles discussed in this contribution. In particular, it would be particularly interesting to extend the THM to cases involving three-fragment projectiles, where three types of reactions can be extracted, $x_1 + A \rightarrow y_1 + B_1$, $x_2 + A \rightarrow y_2 + B_2$, and $x_1+x_2 + A \rightarrow y_3 + B_3$. Work along these lines is in progress.

\section{Conclusions and Perspectives}

We have reported in this contribution that the general structure of the CFH cross section in the DWBA limit is similar in structure to the 3B one
, with the full post form (or four-body IAV), which can be written as the sum of the prior four-body 
UT cross section plus the four-body HM one plus an interference term when the 4B 
source term detailed in Eq. (\ref{CFH}) is used to compute the inelastic part of the inclusive cross-section.
 The major difference between the 4B and 3B 
cases resides in the structure of the reaction cross sections for the absorption of one of the interacting fragments, which 
we find to be damped by the absorption effect of the other fragment.
 It is expected that in a (t,p) reaction, an absorption cross section of the $n + A$ subsystem would be smaller than the corresponding one in a (d,p) reaction.
 Another important new feature is the 3B absorption from a three-body optical potential, the formal structure of which we have sketched. 
 
 The perspectives of our study are in the use of the 
 Faddeev-Yakubovski equations \cite{Faddeev1961, Yakubov1967} to develop an expansion of the four-body wave function, to attempt to
 go beyond the CFH/IAV formulas. This effort could also be accomplished by building the four-body CDCC wave function
 to compute the source term. In addition, we can ask about the dynamics that are built in the three-body optical potential
when giant pairing vibrations (GPV) are possible \cite{Broglia1977,Cappuzz2015}.  These collective states involve the coherent excitation of 
particle-particle pairs, in complete analogy to the coherent excitation of particle-hole pairs that constitutes the microscopic foundation of multipole giant resonances. For example, the GPV opens interesting prospects for nuclear structure studies in reactions of the type (t, p),  2n Borromean cases such as ($^{6}$He, $^4$He), ($^{11}$Li, $^9$Li), ($^{14}$Be, $^{12}$Be), ($^{22}$C, $^{20}$C), and the 2p halo cases, ($^{17}$Ne, $^{15}$O), and
($^{20}$Mg, $^{18}$Ne). The theory we have developed is an appropriate framework  to study these types 
of collective nuclear excitations, associated with pairing correlations in the target.

From the point of view of a two-neutron s-wave dominated halo target our theory naturally leads to an inquiry about the r\"{o}le of
Efimov physics in inclusive breakup cross-section, both in the elastic and inelastic contributions, 
 which should have its place beyond the CFH formula.
  
 Further, the theory can be extended to allow for inclusive reactions where two or more  fragments are
detected. We expect that the  developments discussed in our recent work 
should stimulate  experimental and theoretical works to seek more information about 
the $x_{1} + x_{2}$ system in the elastic and inelastic breakup cross sections, and in particular 
to theorists to extend the Surrogate Method and the Trojan Horse Method, both based on the inclusive non-elastic breakup part 
of the $b$ spectrum in a two-fragment projectile induced reaction, to three-fragment projectiles. \\

{\it Acknowledgements.} 
This work was partly supported by the Brazilian agencies, Funda\c c\~ao de Amparo \`a Pesquisa do Estado de
 S\~ao Paulo (FAPESP), the 
Conselho Nacional de Desenvolvimento Cient\'ifico e Tecnol\'ogico  (CNPq). MSH also acknowledges a Senior Visiting Professorship granted by the Coordena\c c\~ao de Aperfei\c coamento de Pessoal de N\'ivel Superior (CAPES), through the CAPES/ITA-PVS program.


\begin{thebibliography}{99}

\bibitem{CarPLB2017} 	
B. V. Carlson, T. Frederico, M. S. Hussein, Phys. Lett. B \textbf{767}, 53 (2017).

\bibitem{IAV1985}M. Ichimura, N. Austern, and C. M. Vincent, Phys. Rev. C \textbf{32}, 431 (1985).

\bibitem{UT1981}T. Udagawa and T. Tamura, Phys. Rev. C \textbf{24}, 1348 (1981).

\bibitem{HM1985}M. S. Hussein and K. W. McVoy, Nucl. Phys. A \textbf{445}, 124 (1985)

\bibitem{Austern1987}N. Austern, Y. Iseri, M. Kamimura, M. Kawai, G. Rawitscher,
and M. Yahiro, Phys. Rep. \textbf{154}, 125 (1987). 

\bibitem{HFM1990}M. S. Hussein, T. Frederico, and R. C. Mastroleo, Nucl. Phys. A \textbf{511},
269 (1990).

\bibitem{Efimov} V. Efimov, Phys. Lett. B \textbf{33}, 563 (1970) 563; Few-Body Syst. \textbf{51}, 79 (2011).
\bibitem{FrePPNP12}  T. Frederico, M.T. Yamashita, A. Delfino, L. Tomio, Prog. Part. Nucl. Phys. \textbf{67}, 939
(2012).

\bibitem{DDCH2015} P. Descouvemont, T. Druet, L. F. Canto, and M. S. Hussein
Phys. Rev. C \textbf{91}, 024606 (2015).
\bibitem{CH2013} L. F. Canto and M. S. Hussein, \textit{Scattering Theory of Molecules, Atoms and Nuclei}, World Scientific , 2013

\bibitem{Tostevin}P. G. Hansen and J. A. Tostevin, Annu. Rev. Nucl. Part.
Sci. \textbf{53}, 219 (2003).
\bibitem{Bertsch}H. Esbensen and G. F. Bertsch
Phys. Rev. C \textbf{64}, 014608 ( 2001)
\bibitem{MaccFBS15} A. O. Macchiavelli, Few-Body Syst. \textbf{56}, 773 (2015)

\bibitem{InnsNP06}  T. Kraemer, et al., Nature \textbf{440}, 315 (2006).
\bibitem{Faddeev1961} L.D.Faddeev, Zh. Eksp. Teor. Fiz. \textbf{39}, 1459 (1960) [Sov. Phys. JETP \textbf{12}, 1014 (1961)].

\bibitem{Yakubov1967} O. A. Yakubovsky, Yad. Fiz. \textbf{5}, 1312 (1967) [Sov. J. Nucl. Phys. \textbf{5}, 1312 (1967)].


\bibitem{Broglia1977} R. A. Broglia and D. R. Bes, Phys. Lett. B \textbf{69}, 129 (1977).

\bibitem{Cappuzz2015} F. Cappuzzello, D. Carbone, M. Cavallaro, M. Bond, C. Agodi, F. Azaiez, A. Bonaccorso, A. Cunsolo, L. Fortunato, A. Foti, S. Franchoo, E. Khan, R. Linares, J. Lubian, J. A. Scarpaci and A. Vitturi, Nature Communications \textbf{6},  6743 (2015).

\bibitem{Ducasse2015} Q. Ducasse et al., arXiv:1512.06334 [nucl-ex]

\bibitem{Potel2015}G. Potel, F. M. Nunes, and I. J. Thompson, Phys. Rev. C \textbf{92}, 034611 (2015).


\bibitem{Moro2015} J.  Lei and A. M. Moro, Phys. Rev. C \textbf{92}, 044616 (2015);
J. Lei and A. M. Moro, C \textbf{92}, 061602(R) (2015).
\bibitem{Potel2017} G. Potel et al. To be published in the European Physical Journal A (2017)

 	
\bibitem{Carlson2015} B.V. Carlson, R. Capote, M. Sin, Few-Body Syst. {\textbf 57}, 307 
(2016).

\bibitem{SM0}Jutta E. Escher and Frank S. Dietrich, Phys. Rev. C \textbf{74}, 054601 (2006)
\bibitem{SM1} Jutta E. Escher, Jason T. Burke, Frank S. Dietrich, Nicholas D. Scielzo,
Ian J. Thompson, and Walid Younes, Rev. Mod. Phys., \textbf{84}, 353 (2012)
\bibitem{SM2}J. E. Escher, A. P. Tonchev, J. T. Burke, P. Bedrossian, R. J. Casperson, N. Cooper,, R.
O. Hughes, P. Humby, R. S. Ilieva,, S. Ota1,, N. Pietralla,, N. D. Scielzo, and V.Werner, EPJ Web of Conferences \textbf{122}, 12001 (2016)
\bibitem{THM0} S. Typel and G. Baur, Ann. Phys. (NY)  \textbf{305},  228 (2003)
\bibitem{THM1}C. Spitaleri et al., Phys. of Atomic Nuclei \textbf{74}, 1725 (2011) 
\bibitem{THM2} A. Tumino et al., Few Body Systems, (2012) DOI 10.1007/s00601-012-0407-1
\bibitem{THM3} R. Tribble et al., Rep. Prog. Pays. \textbf{77}, 106901(2014) 



\end{thebibliography}
\end{document}